\documentstyle[graphicx,12pt]{article}
\begin{document}

\small
\hoffset=-1truecm
\voffset=-2truecm
\title{\bf The further estimations of the Q-balls
with one-loop motivated effective potential}
\author {Yue Zhong \hspace {1cm}Hongbo Cheng\footnote {E-mail address:
hbcheng@ecust.edu.cn}\\
Department of Physics, East China University of Science and
Technology,\\ Shanghai 200237, China}

\date{}
\maketitle

\begin{abstract}
The analytical estimations on the Friedberg-Lee-Sirlin typed
Q-balls is performed. The two-field Q-balls are also discussed
under the one-loop motivated effective potential subject to the
temperature. We argue under the analytical consideration that the
parameters from the potential can be regulated to lead the energy
per unit charge of Q-balls to be lower to keep the model stable.
If the energy density is low enough, the Q-balls can become
candidates of dark matter. It is also shown rigorously that the
two-field Q-balls can generate in the first-order phase transition
and survive while they are affected by the expansion of the
universe. The analytical evaluations show that the Q-balls with
one-loop motivated effective potential can exist with the
adjustment of coefficients of terms. We cancel the infinity in the
energy to obtain the necessary conditions consist with those
imposed in the previous work. According to the approximate
expressions instead of curves versus the model parameters with a
series of fixed values, the lower temperature will reduce the
energy density, so there probably have been more and more stable
Friedberg-Lee-Sirlin typed Q-balls to become the dark matter in
the expansion of the universe.
\end{abstract}

\vspace{4cm} \hspace{1cm} PACS number(s): 98.80.-k, 14.80.-j,
95.35.+d, 98.80.Ft

\newpage

\noindent \textbf{I.\hspace{0.4cm}Introduction}

In contrast to the topological charge resulting from spontaneously
symmetry breaking in the course of topological defects, the
nontopological solitons involve a conserved Noether charge because
of a symmetry of this kind of Lagrangian system [1-4]. The Q-balls
as nontopological solitons appear in extended localized solutions
of models with a certain self-interacting complex scalar field
[5]. The charge of Q-balls and that their mass is smaller than the
mass of a collection of scalar fields keep them stable instead of
dispersing. The Q-balls are important models which have been
studied in many areas of physics. The Q-balls can become boson
stars as flat spacetime limits [6-8]. We studied the
nontopological solitons in de Sitter and anti de Sitter spacetimes
respectively to show the constrains from background on the models
[9, 10]. The compact Q-balls in the complex Signum-Gordon model
were also explored [11-14]. The Affleck-Dine field fragments into
Q-balls which generated in the early universe and change the
scenario of Affleck-Dine baryogenesis significantly [15-18]. It
should be pointed out that the Q-balls may be studied to explain
the common origin of the baryon asymmetry and the dark matter. The
Affleck-Dine mechanism produces a scalar field condensating with
baryon and the baryon asymmetry while these kinds of
nontopological solitons may be considered as candidates of dark
matter [16-20]. Further the scalar field configuration of the
Q-balls with a step function was discussed to calculate the ratio
of the Q-ball decay into the candidates for dark matter [21-23].
There are more descriptions on the detections of Q-balls [24-27].
In the cosmological context, the existence of the Q-balls was
formulated and further estimate the net baryon number of the
universe, its dark matter and the ratio of the baryon to cold dark
matter [28]. The evolution of universe is associated with change
of temperature. A new kind of first-order phase transition was
induced by that the Q-balls build up quickly with absorbing
charged particles from the outside in the process of expanding
universe with the sufficiently low temperature [29]. There may
exist the phase transitions induced the solitosynthesis that will
lead the formation of large Q-balls in the process of graduate
charge accretion if some primordial charge asymmetry and initial
seed-like Q-balls exist [29-31]. The Q-balls subject to the
thermal logarithmic potential were investigated [32-35].

It is fundamental to have a better understanding of the relations
of the cosmological phase transition, dark matter Q-balls and the
baryon asymmetry. It is interesting that the baryon asymmetry and
dark matter in the universe may have common origin [36-45]. In
this case the production of the baryon asymmetry and the dark
matter may happen. In order to explain the formation of dark
matter, a mechanism that some kinds of Q-balls packing the
particles were put forward [46]. It is further shown that these
kinds of Q-balls generated in the course of the first-order
cosmological phase transition [46, 47]. The Friedberg-Lee-Sirlin
type Q-balls, one of the simplest models [1, 48-50], confine the
particles to become dark matters and the formulae for the dark
matter properties such as charge, mass and concentration of dark
matter Q-balls are obtained [46]. The effective potential of a
particular one-loop motivated form was also introduced to describe
the generation of the dark matter Q-balls during the cosmological
phase transition that is strongly first order to relate the dark
matter Q-ball parameters and their present mass density to the
properties of the first-order phase transition [46]. The phase
transition temperature has also been estimated reasonably [46]. As
the temperature decreased, these Q-balls as candidates of dark
matter shrink, so the Q-ball size is very small [46].

It is important to explore the Friedberg-Lee-Sirlin Q-balls with
an effective potential of one-loop motivated form with the help of
virvial theorem in order to describe the models further. This kind
of Friedberg-Lee-Sirlin Q-balls as the candidate of dark matter
forming in the first-order cosmological phase transition at some
temperature $T_{c}$ and evolving in the expanding universe has
been estimated numerically [46]. It is significant to find the
analytical expressions for charge, radius and energy of some
special Q-balls in order to exhibit the formation and evolution of
this kind of dark matter Q-balls in detail and more accurately. It
is difficult to reveal the reliable and explicit relations among
the model parameters and temperature by performing the burden
numerical calculation repeatedly because the field equations for
the two kinds of fields consisting of Q-balls are nonlinear and
certainly complicated. It was found that a generalized virvial
relation for Q-balls with general potential in the spacetime with
arbitrary dimensionality was used to derive the analytical
description for Q-balls instead of a series of curves [51]. It
should be pointed out that only the analytical expressions can
show clearly how the potential involving the model variables
influences on the existence and the stability of the Q-balls. Here
we will follow the procedure of Ref. [51] to discuss the
Friedberg-Lee-Sirlin type Q-balls with particular field
configuration governed by an effective potential of one-loop
motivated form. We hope to understand the necessary conditions for
the generation of dark matter Q-balls during the cosmological
phase transition and how the model parameters and temperature
influence on the Q-balls and the fate of these kinds of dark
matter.

We describe the special types of dark matter Q-balls with one-loop
motivated effective potential at finite temperature analytically
by means of virvial theorem in this paper. We look for the virvial
relation for the Friedberg-Lee-Sirlin type Q-balls in the case of
zero temperature or nonzero ones respectively. With the help of
corresponding ansatz, we hope to find the radius and energy of
these Q-balls as functions of model parameters and temperature to
show the relation between their existence and the evolution of the
universe. Finally we list our results.

\vspace{0.8cm} \noindent \textbf{II.\hspace{0.4cm}The virial
relation for the Friedberg-Lee-Sirlin type Q-balls}

The Lagrangian of the Friedberg-Lee-Sirlin Q-balls as one of the
simplest compact objects with a global symmetry and associated
conserved charge reads [46, 48-50],

\begin{equation}
{\mathcal{L}}=\frac{1}{2}\partial_{\mu}\phi\partial^{\mu}\phi
+(\partial_{\mu}X)(\partial^{\mu}X)-h^{2}\phi^{2}X^{+}X-U(\phi)
\end{equation}

\noindent with the index $\mu=0, 1, 2, 3$ and the signature $(+,
-, -, -)$. The potential is assumed to be,

\begin{equation}
U(\phi)=\lambda(\phi^{2}-v^{2})^{2}
\end{equation}

\noindent and here the parameters $h$, $\lambda$ and $v$ belong to
the model. The complex scalar field $X=X(x)$ is used to constitute
the Q-ball. The real scalar field $\phi=\phi(\mathbf{r})$ makes up
the external influence. There exist the conditions to keep the
Q-ball stable, so the Lagrangian of the Q-ball has a conserved
$U(1)$ symmetry under the global transformation
$X(x)\longrightarrow e^{i\alpha}X(x)$ with constant $\alpha$ [2].
The associated conserved current density is defined as
$j^{\mu}\equiv-i(X^{+}\partial^{\mu}X-X\partial^{\mu}X^{+})$ and
the corresponding conserved charge can be given by $Q=\int
d^{3}xj^{0}$ [2]. The total potential of the system has a global
minimum at $X=0$ and $\phi=v$ outside the ball. The ansatz for
field configuration with lowest energy is chosen as [51],

\begin{equation}
X(x)=\frac{1}{\sqrt{2}}F(\mathbf{r})e^{i\omega t}
\end{equation}

\noindent Here the fields $F(\mathbf{r})$ and $\phi(\mathbf{r})$
can be taken to be spherically symmetry meaning
$F(\mathbf{r})=F(r)$ and $\phi(\mathbf{r})=\phi(r)$.
$\{\mathbf{r}\}$ represent the spatial components and
$r=|\mathbf{r}|$ is the radial part. The field equations for this
Q-ball read,

\begin{equation}
(\nabla^{2}+\omega^{2})F-h^{2}\phi^{2}=0
\end{equation}

\noindent and

\begin{equation}
\nabla^{2}\phi-h^{2}F^{2}\phi-4\lambda(\phi^{2}-v^{2})\phi=0
\end{equation}

\noindent The Lagrangian (1) can lead the total energy of the
system,

\begin{equation}
E[F, \phi]=\int d^{3}x[\frac{1}{2}(\nabla
F)^{2}+\frac{1}{2}(\nabla\phi)^{2}+\frac{1}{2}\omega^{2}F^{2}+V(F^{2},
\phi^{2})]
\end{equation}

\noindent where the reduced potential is,

\begin{equation}
V(F^{2},
\phi^{2})=\frac{1}{2}h^{2}F^{2}\phi^{2}+\lambda(\phi^{2}-v^{2})^{2}
\end{equation}

\noindent According to Ref. [51], the virial relation thought as a
generalization of Derrick's theorem for Q-balls can be expressed
as,

\begin{equation}
\langle
U(\phi)+h^{2}\phi^{2}X^{+}X\rangle=\frac{1}{2}\frac{Q^{2}}{\langle
F^{2}\rangle}-\frac{1}{6}\langle(\nabla
F)^{2}\rangle-\frac{1}{6}\langle(\nabla\phi)^{2}\rangle
\end{equation}

\noindent where $\langle\cdot\cdot\cdot\rangle=\int\cdot\cdot\cdot
d^{3}x$. The total charge of the Q-ball is,

\begin{eqnarray}
Q=\int j^{0}d^{3}x\nonumber\\
=\omega\langle F^{2}\rangle\hspace{1cm}
\end{eqnarray}

\noindent Because of $\langle
U(\phi)+h^{2}\phi^{2}X^{+}X\rangle\geq0$, the absolute lower bound
for Q-balls to be a preferred energy state becomes,

\begin{equation}
Q^{2}\geq\frac{1}{3}(\langle(\nabla
F)^{2}\rangle+\langle(\nabla\phi)^{2}\rangle)\langle F^{2}\rangle
\end{equation}

\noindent The energy per unit charge is shown as,

\begin{equation}
\frac{E}{Q}=\omega(1+\frac{\langle(\nabla
F)^{2}\rangle+\langle(\nabla\phi)^{2}\rangle}{6\langle
U(\phi)+h^{2}\phi^{2}F^{2}\rangle+\langle(\nabla
F)^{2}\rangle+\langle(\nabla\phi)^{2}\rangle})
\end{equation}

\noindent The necessary conditions can be imposed in Eq. (11) to
keep the Q-balls'stability.

Now we start to study the large Friedberg-Lee-Sirlin Q-balls.
First of all we assume more than one kind of field according to
the Coleman issue [5]. We choose both the complex scalar field $X$
composing the Q-balls and the real scalar field $\phi$ as external
limitation to be step functions. The step function for the complex
field is equal to be a constant $F_{c}$ within the model and
vanishes outside the ball's volume $V_{c}$. In contrast to the
matter field of the Q-balls, we construct the step function for
the real one as zero inside the Q-ball and the parameter $v$
beyond the distribution of ball matter. This kind of step
functions as the field solutions suggest that the system energy
distributes within the ball size. According to the profiles of the
fields for Q-balls above, the system energy is,

\begin{equation}
E=\frac{1}{2}\frac{Q^{2}}{F_{c}^{2}V_{c}}+U(0)V_{c}
\end{equation}

\noindent where $U(0)=U(\phi)|_{\phi=0}=\lambda v^{4}$. By means
of extremizing the expression (12) with respect to the volume
$V_{c}$, the minimum energy per unit charge is,

\begin{eqnarray}
\frac{E_{min}}{Q}=\frac{\sqrt{2}}{F_{c}}\sqrt{U(0)}\nonumber\\
<m\hspace{2.5cm}
\end{eqnarray}

\noindent while the expression can prevent the Q-ball from
dispersing. The Friedberg-Lee-Sirlin type Q-balls within the frame
of Coleman approach can survive.

In order to try to describe the large Friedberg-Lee-Sirlin type
Q-balls, we choose the field profile [51, 52],

\begin{eqnarray}
F(r)=\{\begin{array}{cc}
  F_{c} & r<R \\
  F_{c}e^{-\alpha(r-R)} & r\geq R \\
\end{array}
\end{eqnarray}

\noindent and

\begin{eqnarray}
\phi(r)=\{\begin{array}{cc}
  0 & r<R \\
  v(1-e^{-\alpha(r-R)}) & r\geq R \\
\end{array}
\end{eqnarray}

\noindent where $\alpha$ is a variational parameter and $R$
represents the radius of Q-ball. The choice of the same parameter
$\alpha$ is acceptable. It should be pointed out that the
so-called ansatz (14) and (15) are just assumptions keeping the
Q-balls boundary conditions. We make use of the field
configuration to probe this kind of Q-balls. According to the
large-Q-ball ansatz (14) and (15), the energy of model reads,

\begin{eqnarray}
E[F,
\phi]=\frac{1}{2}\frac{Q^{2}}{F^{2}}+\frac{1}{2}\langle(\nabla
F)^{2}\rangle+\frac{1}{2}\langle(\nabla\phi)^{2}\rangle+\langle
U(\phi)+h^{2}X^{+}X\phi^{2}\rangle\nonumber\\
=\frac{1}{2}\frac{Q^{2}}{\frac{4\pi}{3}F_{c}^{2}R^{3}+8\pi
F_{c}^{2}[\frac{1}{(2\alpha)^{3}}+\frac{R}{(2\alpha)^{2}}
+\frac{R^{2}}{4\alpha}]}\hspace{4cm}\nonumber\\
+\pi\alpha v^{2}[\frac{1}{2\alpha^{2}}+\frac{R}{\alpha}+R^{2}]
+\pi\alpha
F_{c}^{2}[\frac{1}{2\alpha^{2}}+\frac{R}{\alpha}+R^{2}]\hspace{2cm}\nonumber\\
+\frac{4\pi}{3}\lambda v^{4}R^{3}+\pi\lambda
v^{4}(\frac{635}{216\alpha^{3}}+\frac{89R}{18\alpha^{2}}
+\frac{11R^{2}}{3\alpha})\hspace{2.5cm}\nonumber\\
+\pi
h^{2}v^{2}F_{c}^{2}(\frac{115}{432\alpha^{3}}+\frac{13R}{36\alpha^{2}}
+\frac{R^{2}}{6\alpha})\hspace{4cm}
\end{eqnarray}

\noindent where the total conserved charge is,

\begin{equation}
Q=\frac{4\pi}{3}\omega F_{c}^{2}R^{3}+2\pi\omega
F_{c}^{2}(\frac{1}{2\alpha^{3}}+\frac{R}{\alpha^{2}}+\frac{R^{2}}{\alpha})
\end{equation}

\noindent We should further investigate the properties of
Friedberg-Lee-Sirlin Q-balls such as their stability. We extremize
the energy expression (16) with respect to radius $R$ and
coefficient $\alpha$ respectively because this kind of Q-ball
model does not involve the variables $R$ and $\alpha$ which should
be confirmed. We perform the derivation $\frac{\partial
E}{\partial R}|_{R=R_{cl}}=0$ to find the approximate critical
radius of Q-balls,

\begin{equation}
R_{cl}\approx(\frac{9Q^{2}}{32\pi^{2}\lambda
v^{4}F_{c}^{2}})^{\frac{1}{6}}-(\frac{1}{12\pi\lambda
v^{4}})(\alpha v^{2}+\alpha F_{c}^{2}+\frac{11\lambda
v^{4}}{3\alpha}+\frac{h^{2}v^{2}F_{c}^{2}}{6\alpha})
\end{equation}

\noindent where we keep several dominant terms in the expression
of the energy for simplicity and this approximation is acceptable
for large Q-balls. The enormous amount of Q-ball charge can keep
the radius positive while huge. According to the condition
$\frac{\partial E}{\partial\alpha}|_{\alpha=\alpha_{c}}=0$
imposing to the energy, we obtain,

\begin{equation}
\alpha_{c}^{2}=\frac{22\lambda
v^{4}+h^{2}v^{2}F_{c}^{2}}{6(v^{2}+F_{c}^{2})}
\end{equation}

\noindent It is obvious that $\alpha_{c}^{2}$ denoted in Eq. (19)
is positive, which keeps $\alpha_{c}$ real, or the model
consisting of fields $X(x)$ and $\phi(x)$ can not constitute the
Q-balls according to the ansatz (14) and (15). Combining Eq. (18)
and Eq. (19), we find the minimum energy of large
Friedberg-Lee-Sirlin Q-ball per unit charge,

\begin{eqnarray}
E_{min}=E|_{R=R_{cl}, \alpha=\alpha_{c}}\hspace{0.5cm}\nonumber\\
\approx\frac{\sqrt{2}\lambda^{\frac{1}{2}}v^{2}}{F_{c}}Q(1+\xi_{c}Q^{-\frac{1}{3}})
\end{eqnarray}

\noindent where

\begin{equation}
\xi_{c}=\frac{3^{\frac{1}{6}}\pi^{\frac{1}{3}}}{2^{\frac{5}{3}}}
\frac{F_{c}^{\frac{1}{3}}(v^{2}+F_{c}^{2})^{\frac{1}{2}}(22\lambda
v^{2}+h^{2}F_{c}^{2})^{\frac{1}{2}}}{\lambda^{\frac{5}{6}}v^{\frac{7}{3}}}
\end{equation}

\noindent The $\xi_{c}$-term in Eq. (20) is due to the dominant
terms. This term is tiny compared to the first one. In the case of
huge Q-ball, its minimum energy per unit charge is,

\begin{equation}
\lim_{Q\longrightarrow\infty}\frac{E[F_{c}, v]|_{R=R_{cl,
\alpha=\alpha_{c}}}}{Q}=\frac{\sqrt{2}\lambda^{\frac{1}{2}}v^{2}}{F_{c}}
\end{equation}

\noindent The explicit expression above shows that the minimum
energy over total charge for this kind of large Q-balls is finite
even the number of particles is extremely large. We can choose the
values of $\lambda$, $v$ and $F_{c}$ to keep the particles of the
system to distribute within a definite region which is lower than
the necessary kinetic energy of a free particle in the system, and
this kind of Q-balls keep stable. The Eq. (14) and Eq. (15) show
that the field $F(r)$ decreases while the field $\phi(r)$
increases as $r>R$. When $r$ as Q-ball size is large enough, the
fields $F(r)$ and $\phi(r)$ approach to the zero and $v$
respectively, leading the total energy (6) with the reduced
potential (7) to vanish and satisfying the boundary conditions.
The numerical solutions to field equations (4) and (5) satisfy the
requirements of $F(r)$ and $\phi(r)$ for large
Friedberg-Lee-Sirlin Q-ball in the Figure 1. The minimum energy
density certainly depends on the model parameters such as $h$,
$\lambda$ and $v$. The influence from the variable $\lambda$ on
the relation between the minimum energy per unit charge of the
Friedberg-Lee-Sirlin Q-ball and the charge amount $Q$ is depicted
in the Figure 2. The variable $\lambda$ with larger magnitude will
increase the energy. The corrections from the other parameters $h$
and $v$ are similar to the results above. The solutions
corresponding to the ansatz (14) and (15) can compose the Q-balls.

Now we pay attention to the small Friedberg-Lee-Sirlin type
Q-balls. According to Ref. [51, 52], we also bring about a
Gaussian ansatz like,

\begin{equation}
F(r)=F_{c}e^{-\frac{r^{2}}{R^{2}}}
\end{equation}

\noindent and

\begin{equation}
\phi(r)=v(1-e^{-\frac{r^{2}}{R^{2}}})
\end{equation}

\noindent to give an example. We explore the special Q-balls whose
field profiles obey the ansatz (23) and (24). We substitute the
ansatz (23) and (24) into the expression (6) to find the total
energy belonging to the small Q-balls as follows,

\begin{eqnarray}
E=\int d^{3}x[\frac{1}{2}\omega^{2}F^{2}+\frac{1}{2}(\nabla
F)^{2}+\frac{1}{2}(\nabla\phi)^{2}+U(\phi)+\frac{1}{2}h^{2}\phi^{2}F^{2}]\nonumber\\
=aR^{-3}+bR+cR^{3}\hspace{7cm}
\end{eqnarray}

\noindent where the charge is,

\begin{equation}
Q=(\frac{\pi}{2})^{\frac{3}{2}}\omega F_{c}^{2}R^{3}
\end{equation}

\noindent replacing the frequency $\omega$. In the process similar
to that of Ref. [51], we extremize the expression of the energy
with respect to $R$ like $\frac{\partial E}{\partial
R}|_{R=R_{cs}}=0$ to set up the equation for the critical radius
$R_{cs}$ as follows,

\begin{equation}
3cR_{cs}^{6}+bR_{cs}^{4}-3a=0
\end{equation}

\noindent where

\begin{equation}
a=\frac{2^{\frac{1}{2}}Q^{2}}{\pi^{\frac{3}{2}}F_{c}^{2}}
\end{equation}

\begin{equation}
b=\frac{3}{2}(\frac{\pi}{2})^{\frac{3}{2}}(F_{c}^{2}+v^{2})
\end{equation}

\begin{equation}
c=(\frac{1}{8}-4\times
3^{-\frac{3}{2}}+2^{\frac{1}{2}})\pi^{\frac{3}{2}}\lambda
v^{4}+(2^{-\frac{5}{2}}-3^{-\frac{3}{2}}
+\frac{1}{16})\pi^{\frac{3}{2}}h^{2}v^{2}F_{c}^{2}
\end{equation}

\noindent The acceptable approximate solution for special critical
radius becomes,

\begin{equation}
R_{cs}=[1-\frac{b}{2(9cR_{0}^{2}+2b)}]R_{0}
\end{equation}

\noindent with

\begin{equation}
R_{0}=(\frac{2^{\frac{1}{2}}}{\pi^{\frac{3}{2}}F_{c}^{2}})^{\frac{1}{6}}Q^{\frac{1}{3}}
[(\frac{1}{8}-\frac{4}{3^{\frac{3}{2}}}+\sqrt{2})\pi^{\frac{3}{2}}\lambda
v^{4}+(\frac{1}{2^{\frac{5}{2}}}-\frac{1}{3^{\frac{3}{2}}}+\frac{1}{16})
\pi^{\frac{3}{2}}h^{2}v^{2}F_{c}^{2}]^{-\frac{1}{6}}
\end{equation}

\noindent The minimum energy in the case of small Q-balls can be
shown in terms of the critical radius as,

\begin{eqnarray}
E_{min}=E[F_{c}]|_{R=R_{cs}}\hspace{5cm}\nonumber\\
\approx(\frac{2^{\frac{1}{2}}}{\pi^{\frac{3}{2}}cF_{c}^{2}})^{\frac{1}{2}}
[2c+b(\frac{c}{\tilde{a}})^{\frac{1}{3}}
Q^{-\frac{2}{3}}-\frac{b^{2}}{18}(\tilde{a}^{2}c)^{-\frac{1}{3}}Q^{-\frac{4}{3}}]Q
\end{eqnarray}

\noindent where
$\tilde{a}=\frac{\sqrt{2}}{\pi^{\frac{3}{2}}}\frac{1}{F_{C}^{2}}$.
The amount of charge for smaller balls can also extremely large.
In the limit of huge quantity of charge, the minimum energy per
unit charge for small Q-balls becomes,

\begin{equation}
\lim_{Q\longrightarrow\infty}\frac{E[F_{c}]|_{R=R_{cs}}}{Q}
=2(\frac{2^{\frac{1}{2}}c}{\pi^{\frac{3}{2}}F_{c}^{2}})^{\frac{1}{2}}
\end{equation}

\noindent It is clear that the energy density will not be
divergent. For the Gauss ansatz (23) and (24) [51, 52], the small
Friedberg-Lee-Sirlin typed Q-ball constitutions $F(r)$ and
$\phi(r)$ are decreasing and increasing functions respectively at
first and approach to the constants such as zero and Q-ball
variable $v$. In the case of small balls, the profiles of the
solutions to field equations (4) and (5) shown numerically in the
Figure 3 meet the Q-balls requirements including the necessary
boundary conditions. According to the Figure 4, the shapes of the
curves for small balls energy density resemble those of large
Friedberg-Lee-Sirlin typed Q-balls. The model variable $\lambda$
with larger value can also make the total energy greater. This
kind of small Q-balls can be composed of the solutions subject to
the ansatz (23) and (24). The energy per unit charge can be small
enough if the amount of model variables is selected reasonably. As
an example, these special Q-balls with sufficiently small energy
density can gather the particles instead of decentralizing and
become the dark matter.

\vspace{0.8cm} \noindent \textbf{III.\hspace{0.4cm}The virial
relation for the Friedberg-Lee-Sirlin typed Q-balls with one-loop
motivated effective potential}

We start to investigate the Friedberg-Lee-Sirlin Q-balls subject
to one-loop motivated effective potential involving temperature
from Ref. [46] analytically. Certainly we should choose the
potential to be [46],

\begin{equation}
U(\phi, T)=\alpha'(T^{2}-T_{c_{2}}^{2})\phi^{2}-\gamma
T\phi^{3}+\lambda\phi^{4}
\end{equation}

\noindent where $T_{c_{2}}$, $\alpha'$ and $\gamma$ are model
parameters. With the relation $\alpha' T_{c_{2}}^{2}=2\lambda
v^{2}$, the potential (35) can be thought as the generalization of
potential (2) for the Friedberg-Lee-Sirlin type Q-balls without
thermal corrections like [46],

\begin{equation}
U(\phi, T)=U(\phi)+(-\gamma T\phi^{3}+\alpha'
T^{2}\phi^{2}-\lambda v^{4})
\end{equation}

\noindent We continue our research on this kind of Q-balls in
virtue of technique in Ref. [51]. The total energy from the
Lagrangian of the system is chosen as,

\begin{equation}
E[F, \phi, T]=\frac{1}{2}\omega
Q+\frac{1}{2}\langle(\nabla\phi(r))^{2}\rangle
+\frac{1}{2}\langle(\nabla F(r))^{2}\rangle+\langle U(\phi,
T)\rangle
\end{equation}

We also research on the special Friedberg-Lee-Sirlin models with
ansatz (14), (15), (23) and (24) in the cases of large size and
small ones respectively. At first we substitute the large-Q-ball
ansatz (14) and (15) [51, 52] in Eq. (37) to find that the first
three terms are the same as the parts of Eq. (16). We discuss the
potential terms at finite temperature,

\begin{eqnarray}
\langle U(\phi, T)\rangle=\int U(\phi, T)d^{3}x\hspace{4cm}\nonumber\\
=4\pi v^{2}[\alpha'(T^{2}-T_{c_{2}}^{2})-\gamma Tv+\lambda
v^{2}]\int_{R}^{\infty}r^{2}dr\hspace{1.5cm}\nonumber\\
+8\pi v^{2}\alpha'(T^{2}-T_{c_{2}}^{2})\sum_{k=0}^{2}\frac{1}{k!}
(\frac{1}{2^{3-k}}-2)\frac{R^{k}}{\alpha^{3-k}}\hspace{1.5cm}\nonumber\\
+8\pi v^{3}\gamma T\sum_{k=0}^{2}\frac{1}{k!}(3-\frac{3}{2^{3-k}}
+\frac{1}{3^{3-k}})\frac{R^{k}}{\alpha^{3-k}}\hspace{1.5cm}\nonumber\\
+8\pi
v^{4}\lambda\sum_{k=0}^{2}\frac{1}{k!}(-4+\frac{6}{2^{3-k}}-\frac{4}{3^{3-k}}
+\frac{1}{4^{3-k}})\frac{R^{k}}{\alpha^{3-k}}
\end{eqnarray}

\noindent It is manifest that the integral form
$\int_{R}^{\infty}r^{2}dr$ is equal to the infinity, which leads
the total energy to be divergent. In order to eliminate the
infinity, we choose the coefficient of the integral term to be
zero,

\begin{equation}
\lambda v^{2}-\gamma Tv+\alpha'(T^{2}-T_{c_{2}}^{2})=0
\end{equation}

\noindent like that imposed in Ref. [46]. We have to keep the
following condition,

\begin{equation}
T<T_{c}
\end{equation}

\noindent where the critical temperature is defined as [46],

\begin{equation}
T_{c}^{2}=\frac{4\alpha'\lambda}{4\alpha'\lambda-\gamma^{2}}T_{c_{2}}^{2}
\end{equation}

\noindent to keep the parameter $v$ to be real. The condition (40)
appeared in Ref. [46] when the authors discussed the effective
potential. Here we should pointed out that the necessary condition
prevents the total energy of the system from infinity and leads
the variable $v$ to be real. Having imposed the constraint (39) on
the temperature-corrected potential (38), we obtain the energy of
large Friedberg-Lee-Sirlin type Q-balls containing the temperature
as follows,

\begin{eqnarray}
E[\phi, F]=\frac{1}{2}\omega
Q+\frac{1}{2}\langle(\nabla\phi(r))^{2}\rangle
+\frac{1}{2}\langle(\nabla F(r))^{2}\rangle+\langle
U(\phi)+h^{2}\phi^{2}X^{+}X\rangle\nonumber\\
=\frac{3Q^{2}}{8\pi
F_{c}^{2}}\frac{1}{1+\frac{3}{2}\frac{1}{\alpha
R}}R^{-3}\hspace{8cm}\nonumber\\
+4\pi(\alpha^{2}v^{2}+\alpha^{2}F_{c}^{2}+h^{2}v^{2}F_{c}^{2})
(\frac{1}{8\alpha^{3}}+\frac{R}{4\alpha^{2}}+\frac{R^{2}}{4\alpha})\hspace{3cm}\nonumber\\
+8\pi v^{2}\alpha(T^{2}-T_{c_{2}}^{2})\sum_{k=0}^{2}\frac{1}{k!}
(\frac{1}{2^{3-k}}-2)\frac{R^{k}}{\alpha^{3-k}}\hspace{4cm}\nonumber\\
+8\pi v^{3}\gamma T\sum_{k=0}^{2}\frac{1}{k!}(1-\frac{1}{2^{2-k}}
+\frac{1}{3^{3-k}})\frac{R^{k}}{\alpha^{3-k}}\hspace{4cm}\nonumber\\
+8\pi
v^{4}\lambda\sum_{k=0}^{2}\frac{1}{k!}(-2+\frac{5}{2^{3-k}}-\frac{4}{3^{3-k}}
+\frac{1}{4^{3-k}})\frac{R^{k}}{\alpha^{3-k}}\hspace{2.5cm}\nonumber\\
-4\pi
h^{2}v^{2}F_{c}^{2}(\frac{15}{64}\frac{1}{\alpha^{3}}+\frac{7}{16}\frac{R}{\alpha^{2}}
+\frac{1}{6}\frac{R^{2}}{\alpha})\hspace{4.5cm}
\end{eqnarray}

\noindent According to the similar procedure above [51] and the
reduced energy, the approximate critical radius of Q-ball
associated with the temperature is denoted as,

\begin{equation}
R_{clT}=(\frac{9Q^{2}}{16\pi
v^{2}})^{\frac{1}{5}}[\pi\alpha'(v^{2}+F_{c}^{2})+\frac{4\pi
v^{3}}{3\alpha}\gamma T-\frac{7\pi
v^{4}}{3\alpha}\lambda+\frac{\pi
h^{2}v^{2}F_{c}^{2}}{6\alpha}]^{-\frac{1}{5}}
\end{equation}

\noindent The stable Q-balls expand with decreasing temperature.
With the imposing condition $\frac{\partial
E}{\partial\alpha}|_{\alpha=\alpha_{cT}}=0$, the result is listed
as,

\begin{equation}
\alpha_{cT}^{2}=\frac{v^{2}}{6(v^{2}+F_{c}^{2})}(8v\gamma
T-14v^{2}\lambda+h^{2}F_{c}^{2})
\end{equation}

\noindent and the minimum energy density of large
Friedberg-Lee-Sirlin typed Q-ball at finite temperature is,

\begin{eqnarray}
\frac{E_{min}}{Q}=\frac{E|_{R=R_{clT}, \alpha=\alpha_{cT}}}{Q}
\hspace{2.5cm}\nonumber\\
=[(\frac{2}{3})^{\frac{3}{5}}+(\frac{3}{2})^{\frac{2}{5}}]
[\frac{9\pi(v^{2}+F_{c}^{2})^{3}}{8F_{c}^{4}}]^{\frac{1}{5}}
\alpha_{cT}^{\frac{3}{5}}Q^{-\frac{1}{5}}
\end{eqnarray}

\noindent It should be pointed out that the energy density has
something to do with the charge $Q$ and temperature. For the huge
balls with large amount of charge $Q$, their energy densities will
not be infinite, but to be smaller. It is even more possible for
the huge Q-balls to become the dark matter. According to the Eq.
(44) and Eq. (45), the smaller parameter $\alpha_{cT}$ also leads
the minimum energy density to be smaller. The Figure 5 indicates
that the Q-ball energy will become smaller when the temperature
drops. The expansion of the universe impels more of these kinds of
models to stand for the dark matter because the decreasing
temperature diminishes the parameter $\alpha_{cT}$ to make the
energy densities to be sufficiently low.

Following the same procedure of Ref. [51] with the help of ansatzs
(23) and (24) [51, 52], we have the critical radius of small
Friedberg-Lee-Sirlin typed Q-ball under the one-loop motivated
effective potential as follows,

\begin{equation}
R_{csT}=[1-\frac{b}{2(9c_{T}R_{0T}^{2}+2b)}]R_{0T}
\end{equation}

\noindent where

\begin{equation}
R_{0T}=\frac{2_{\frac{1}{12}}Q^{\frac{1}{3}}}
{\pi^{\frac{1}{2}}v^{\frac{1}{3}}F_{c}^{\frac{1}{3}}}
[(\frac{2-\sqrt{2}}{2}+\frac{\sqrt{3}}{9})\gamma
vT-(\frac{15-10\sqrt{2}}{8}+\frac{4\sqrt{3}}{9})\lambda
v^{2}+(\frac{2\sqrt{2}+1}{16}-\frac{\sqrt{3}}{9})h^{2}F_{c}^{2}]^{-\frac{1}{6}}
\end{equation}

\noindent Further the minimum energy per unit charge of this kind
of Q-balls with small size is,

\begin{eqnarray}
E_{Tmin}=E[F_{c}]|_{R=R_{csT}}\hspace{5cm}\nonumber\\
\approx(\frac{2^{\frac{1}{2}}}{\pi^{\frac{3}{2}}c_{T}F_{c}^{2}})^{\frac{1}{2}}
[2c_{T}+b(\frac{c_{T}}{\tilde{a}})^{\frac{1}{3}}
Q^{-\frac{2}{3}}-\frac{b^{2}}{18}(\tilde{a}^{2}c_{T})^{-\frac{1}{3}}Q^{-\frac{4}{3}}]Q
\end{eqnarray}

\noindent where

\begin{equation}
c_{T}=[(\frac{2-\sqrt{2}}{2}+\frac{\sqrt{3}}{9})\gamma
vT-(\frac{15-10\sqrt{2}}{8}+\frac{4\sqrt{3}}{9})\lambda
v^{2}+(\frac{2\sqrt{2}+1}{16}-\frac{\sqrt{3}}{9})h^{2}F_{c}^{2}]
\pi^{\frac{3}{2}}v^{2}
\end{equation}

\noindent This result (48) is just the replacement of $c$ by
$c_{T}$ in Eq. (33). Also the minimum energy per unit charge for
small one-loop-potential-controlled Q-ball with larger quantity of
charge is,

\begin{equation}
\lim_{Q\longrightarrow\infty}\frac{E[F_{c}]|_{R=R_{cs},
\alpha=\alpha_{cT}}}{Q}
=2(\frac{2^{\frac{1}{2}}c_{T}}{\pi^{\frac{3}{2}}F_{c}^{2}})^{\frac{1}{2}}
\end{equation}

\noindent The energy density relating to the temperature subject
to $c_{T}$ is finite. According to Eq. (49), the parameter $c_{T}$
will become smaller when the temperature is lower. The small
Q-ball energy densities as functions of charge $Q$ for the
temperature are described in the Figure 6. The lower temperature
also leads the energy densities of the small Friedberg-Lee-Sirlin
typed Q-balls with one-loop motivated potential to be smaller. As
the universe expands with the decreasing temperature, the energy
densities of more and more Friedberg-Lee-Sirlin typed Q-balls with
one-loop motivated effective potential will be small enough.
Certainly these small Q-balls can survive and further act as
candidates of dark matter.

\vspace{0.8cm} \noindent \textbf{IV.\hspace{0.4cm}Discussion and
Conclusion}

The Q-balls with two kinds of scalar fields satisfying the special
ansatz produced in the first-order phase transition are considered
analytically. The system of two kinds of fields composing the
Q-balls can become the candidate of dark matter while can also be
used to explain the baryon asymmetry [46]. This kind of Q-balls
are powerful [46] and their existence, stability and evolution
need to be analyzed in detail. It should be emphasized that the
nonlinear field equations of the two kinds of scalar fields
describing the Q-balls are complicated and there are diverse
numerical estimations corresponding to the model constructions
[46]. In the cases of large special balls and relatively small
ones we prove strictly to confirm that the two-field Q-balls will
be stable instead of dispersing if their energy density is smaller
than the kinetic energy per unit charge. Further the Q-balls can
become the dark matter with the sufficiently low energy density.
We study the Friedberg-Lee-Sirlin Q-balls with the appointed
ansatz under the one-loop motivated effective potential with the
burden of derivation. We cancel the divergence in the energy of
the Friedberg-Lee-Sirlin typed Q-ball with one-loop motivated
effective potential and keep the scalar field $\phi$ to be real to
confirm the critical temperature $T_{c}$ and show the existence of
the Q-balls when the temperature $T<T_{c}$ instead of imposing the
conditions based on the numerical calculation like Ref. [46]. Our
explicit expressions for the energy also indicate that the
temperature-dependent Q-balls can act as dark matter when the
energy per unit charge is low enough. The evolution of the Q-balls
in the cosmological background is exhibited without numerical
estimation. The expansion of the universe helps more and more this
kind of Friedberg-Lee-Sirlin type Q-balls under the one-loop
motivated effective potential to become the dark matter because
the decreasing temperature leads the energy per unit charge of the
models to be lower. Here the large or small Friedberg-Lee-Sirlin
typed Q-balls that we study analytically are special because their
fields configurations are particular, but we exhibit their
properties mentioned above and they are consistent with those of
Ref. [46] and of universal significance. The analytical discussion
is reliable and shows the model properties clearly.

\vspace{3cm}

\noindent\textbf{Acknowledgement}

This work is supported by NSFC No. 10875043.

\newpage
\begin{figure}
\setlength{\belowcaptionskip}{10pt} \centering
\includegraphics[width=15cm]{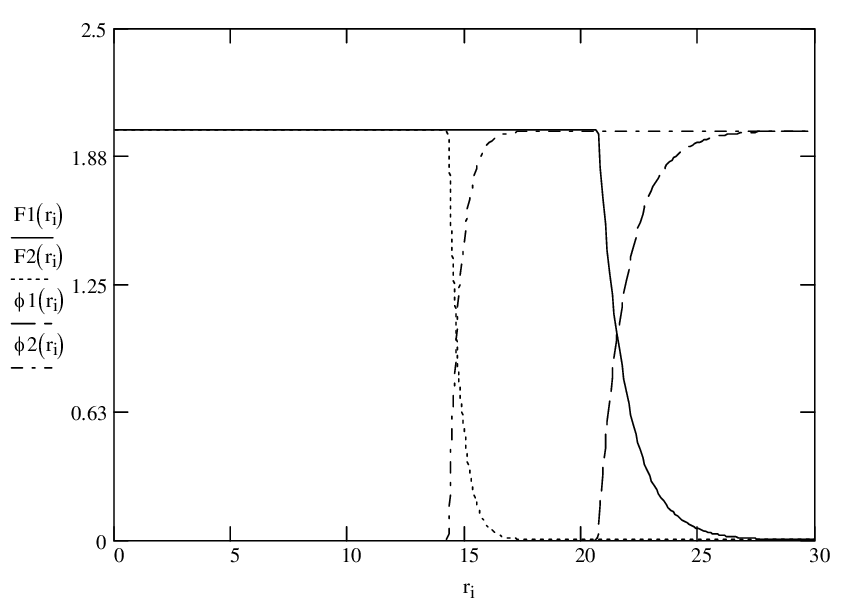}
\caption{The profile function $F(r)$ of large Friedberg-Lee-Sirlin
typed Q-balls with $Q=10^{5}$, $F_{c}=2$, $v=2$ and $h=1$ shown as
solid, dotted curves for coefficient $\lambda=0.05, 0.5$
respectively and the profile function $\phi(r)$ of the same
Q-balls as dashed, dot-dashed curves for coefficient
$\lambda=0.05, 0.5$ respectively.}
\end{figure}

\newpage
\begin{figure}
\setlength{\belowcaptionskip}{10pt} \centering
\includegraphics[width=15cm]{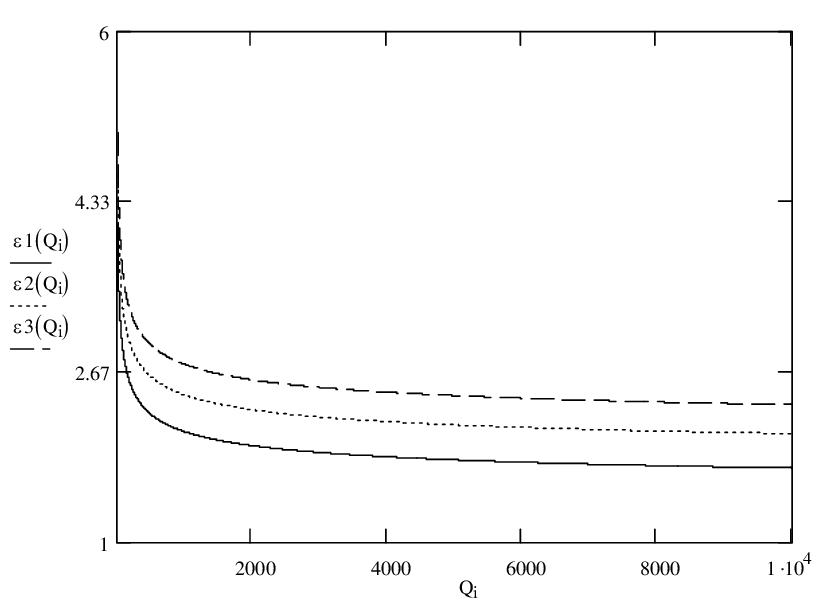}
\caption{The solid, dot, dashed curves of the minimum energy per
unit charge of large Friedberg-Lee-Sirlin typed Q-balls as
functions of charge $Q$ for coefficient $\lambda=2, 4, 6$
respectively.}
\end{figure}

\newpage
\begin{figure}
\setlength{\belowcaptionskip}{10pt} \centering
\includegraphics[width=15cm]{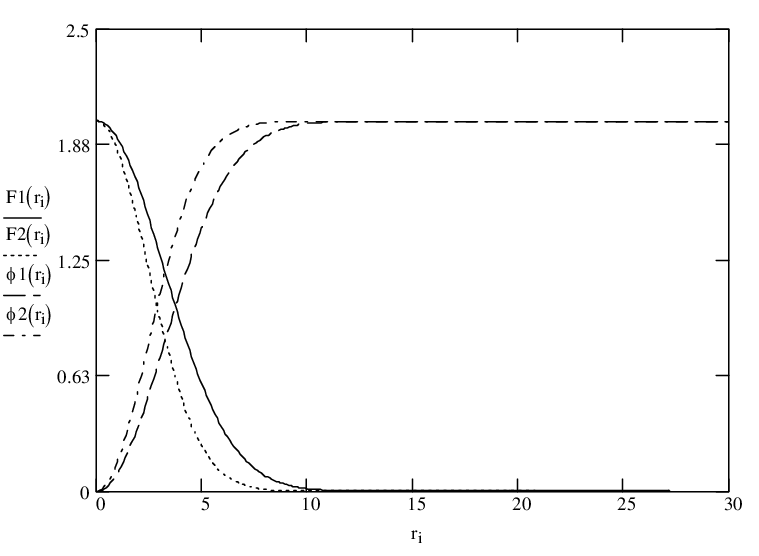}
\caption{The profile function $F(r)$ of small Friedberg-Lee-Sirlin
typed Q-balls with $Q=10^{3}$, $F_{c}=2$, $v=2$ and $h=1$ shown as
solid, dotted curves for coefficient $\lambda=0.05, 0.5$
respectively and the profile function $\phi(r)$ of the same
Q-balls as dashed, dot-dashed curves for coefficient
$\lambda=0.05, 0.5$ respectively.}
\end{figure}

\newpage
\begin{figure}
\setlength{\belowcaptionskip}{10pt} \centering
\includegraphics[width=15cm]{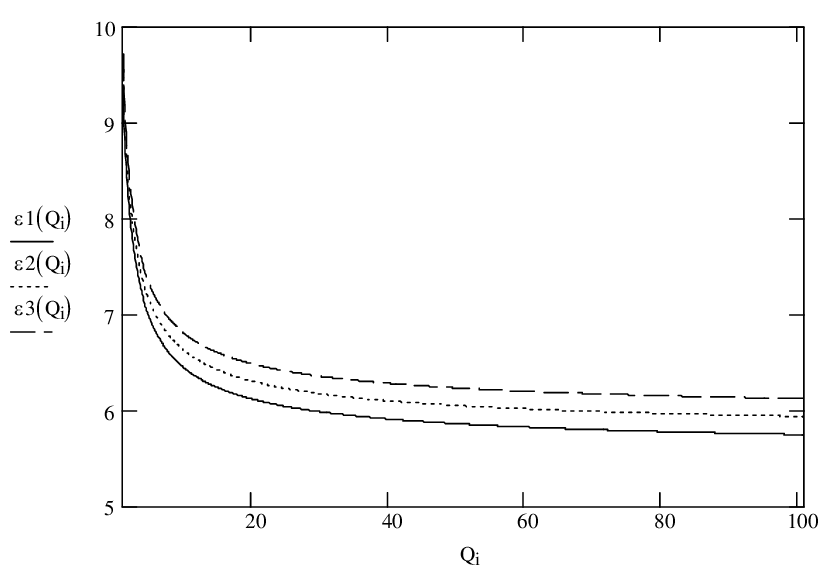}
\caption{The solid, dot, dashed curves of the minimum energy per
unit charge of small Friedberg-Lee-Sirlin typed Q-balls as
functions of charge $Q$ for coefficient $\lambda=2, 4, 6$
respectively.}
\end{figure}

\newpage
\begin{figure}
\setlength{\belowcaptionskip}{10pt} \centering
\includegraphics[width=15cm]{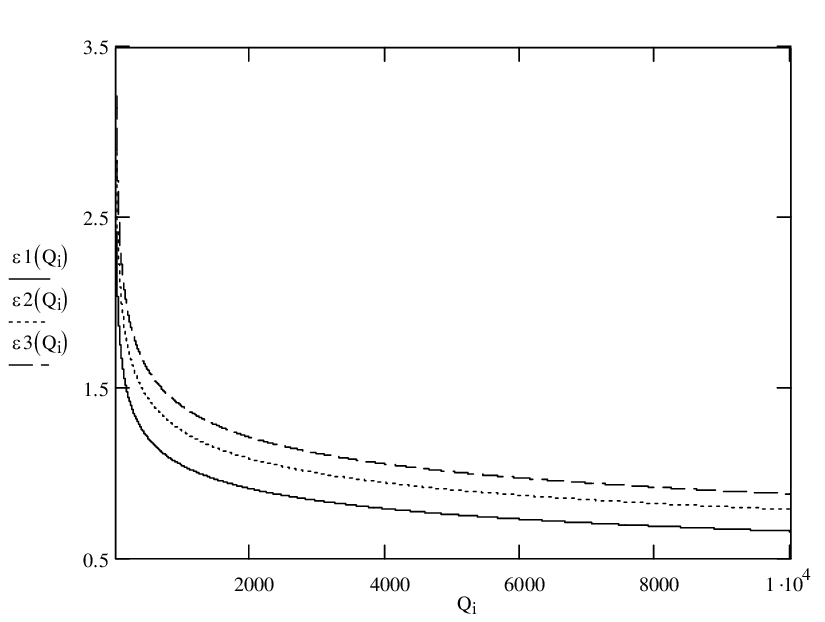}
\caption{The solid, dot, dashed curves of the minimum energy per
unit charge of large Friedberg-Lee-Sirlin typed Q-balls with
one-loop motivated effective potential as functions of charge $Q$
for temperature $T=2, 4, 6$ respectively.}
\end{figure}

\newpage
\begin{figure}
\setlength{\belowcaptionskip}{10pt} \centering
\includegraphics[width=15cm]{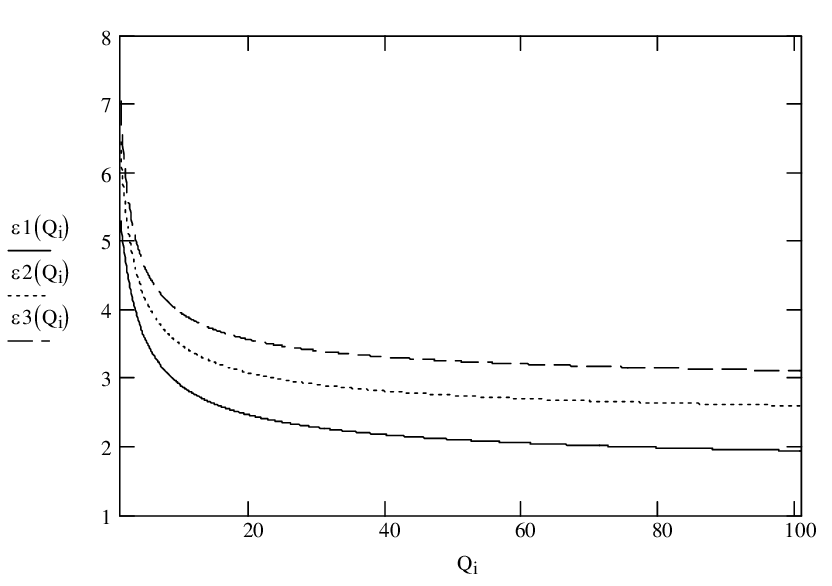}
\caption{The solid, dot, dashed curves of the minimum energy per
unit charge of small Friedberg-Lee-Sirlin typed Q-balls with
one-loop motivated effective potential as functions of charge $Q$
for temperature $T=2, 4, 6$ respectively.}
\end{figure}


\newpage
\begin{thebibliography}{99}
\bibitem{Friedberg}R. Friedberg, T. D. Lee, Y. Pang, Phys.
Rev.D35(1987)3658
\bibitem{Lee}T. D. Lee, Y. Pang, Phys. Rep. 221(1992)251
\bibitem{Enqvist}K. Enqvist, A. Mazumdar, Phys. Rep. 380(2003)99
\bibitem{Dine}M. Dine, A. Kusenko, Rev. Mod. Phys. 76(2004)1
\bibitem{Colemann}S. R. Coleman, Nucl. Phys. B 262(1985)263
\bibitem{Kleihaus}B. Kleihaus, J. Kunz, M. List, Phys. Rev.
D72(2005)064002
\bibitem{Kleihaus}B. Kleihaus, J. Kunz, M. List, I. Schaffer,
Phys. Rev. D77(2008)064025
\bibitem{Brihaye}Y. Brihaye, B. Hartmann, Phys. Rev.
D79(2009)064013
\bibitem{Cheng}H. Cheng, Z. Gu, J. East China Univ. Sci. Technol.
31(2005)509
\bibitem{Cheng}H. Cheng, Z. Gu, J. East China Univ. Sci. Technol.
33(2007)294
\bibitem{Arodz}H. Arodz, P. Klimas, T. Tyranowski, Acta Phys. Pol.
B36(2005)3861
\bibitem{Arodz}H. Arodz, J. Lis, Phys. Rev. D77(2008)107702
\bibitem{Arodz}H. Arodz, J. Lis, Phys. Rev. D79(2009)045002
\bibitem{Wang}H. Wang, H. Cheng, Chin. Phys. Lett. 28(2011)121101
\bibitem{Kawasaki}M. Kawasaki, N. Takeda, JCAP1407(2014)038\\
I. Affeck, M. Dine, Nucl. Phys. B249(1985)361\\
M. Dine, L. Randall, S. D. Thomas, Nucl. Phys. B458(1996)291\\
A. Kusenko, M. E. Shaposhnikov, Phys. Lett. B418(1998)46
\bibitem{Enqvist}K. Enqvist, J. McDonald, Phys. Lett.
B425(1998)309\\
K. Enqvist, J. McDonald, Nucl. Phys. B538(1999)321
\bibitem{Kasuya}S. Kasuya, M. Kawasaki, Phys. Rev. D61(2000)041301
\bibitem{Kasuya}S. Kasuya, M. Kawasaki, Phys. Rev. D62(2000)023512
\bibitem{Kasuya}S. Kasuya, M. Kawasaki, Phys. Rev. D64(2001)123515
\bibitem{Kasuya}S. Kasuya, M. Kawasaki, Phys. Rev. Lett.
85(2000)2677
\bibitem{Kawasaki}S. Kawasaki, M. Yamada, Phys. Rev.
D87(2013)023517
\bibitem{Kasuya}S. Kasuya, M. Kawasaki, M. Yamada,
Phys. Lett. B726(2013)1
\bibitem{Kasuya}S. Kasuya, M. Kawasaki, Phys. Rev. D89(2014)103534
\bibitem{Dvali}G. R. Dvali, A. Kusenko, M. E. Shaposhnikov, Phys.
Lett. B417(1998)99
\bibitem{Kusenko}A. Kusenko, V. Kuzmin, M. E. Shaposhnikov, P. G.
Tinyakov, Phys. Rev. Lett. 80(1998)3185
\bibitem{Cecchini}S. Cecchini et.al., Eur. Phys. J. C57(2008)525
\bibitem{Takenaga}Y. Takenaga, et.al., Phys. Lett. B647(2007)18
\bibitem{Laine}M. Laine, M. Shaposhnikov, Nucl. Phys.
B532(1998)376
\bibitem{Kusenko}A. Kusenko, Phys. Lett. B406(1997)26
\bibitem{Postma}M. Postma, Phys. Rev. D65(2002)085035
\bibitem{Pearce}L. Pearce, Phys. Rev. D85(2012)125022
\bibitem{Kasuya}S. Kasuya, Phys. Rev. D81(2010)083507
\bibitem{Mukaida}K. Mukaida, K. Nakayama, JCAP1301(2013)017
\bibitem{Chiba}T. Chiba, K. Kamada, S. Kasuya, Phys. Rev.
D82(2010)103534
\bibitem{Zhong}Y. Zhong, H. Cheng, Phys. Lett. B743(2015)347
\bibitem{Kitano}R. Kitano, I. Low, Phys. Rev. D71(2005)023510
\bibitem{Farrar}G. R. Farrar, G. Zaharijas, Phys. Rev. Lett.
96(2006)041302
\bibitem{Suematsu}D. Suematsu, JCAP0601(2006)026
\bibitem{An}H. An, S. Chen, R. N. Mohapatra, Y. Zhang,
JHEP1003(2010)124
\bibitem{McDonald}J. McDonald, Phys. Rev. D83(2011)083509
\bibitem{McDonald}J. McDonald, Phys. Rev. D84(2011)103514
\bibitem{Buchmuller}W. Buchmuller, K. Schmitz, G. Vertongen, Nucl.
Phys. B851(2011)481
\bibitem{Poplawski}N. J. Poplawski, Phys. Rev. D83(2011)084033
\bibitem{Davoudiasl}H. Davoudiasl, R. N. Mohapatra, New J. Phys.
14(2012)095011
\bibitem{Barr}S. M. Barr, Phys. Rev. D85(2012)013001
\bibitem{Krylov}E. Krylov, A. Levin, V. Rubakov, Phys. Rev.
D87(2013)083528
\bibitem{Petraki}K. Petraki, M. Trodden, R. R. Volkas,
JCAP1202(2012)044
\bibitem{Friedberg}R. Friedberg, T. D. Lee, A. Sirlin, Phys. Rev.
D13(1976)2739
\bibitem{Friedberg}R. Friedberg, T. D. Lee, A. Sirlin, Nucl. Phys.
B115(1976)1
\bibitem{Friedberg}R. Friedberg, T. D. Lee, A. Sirlin, Nucl. Phys.
B115(1976)32
\bibitem{Gleiser}M. Gleiser, J. Thorarinson, Phys. Rev.
D73(2006)065008
\bibitem{Ioannidou}T. A. Ioannidou, N. D. Vlachos, "On the Q-Ball Profile
Function", hep-th/0302031


\end{thebibliography}
\end{document}